\documentclass[prl,twocolumn,showpacs]{revtex4}
\usepackage{amsmath}
\usepackage{bm}
\usepackage{graphicx}
\usepackage{color}
\usepackage{float}
\usepackage{graphicx}

\begin{document}

\title{Hyperuniform density fluctuations and diverging dynamic correlations 
in periodically driven colloidal suspensions}

\author{Elsen Tjhung}

\author{Ludovic Berthier}

\affiliation{Laboratoire Charles Coulomb, UMR 5221, CNRS and 
Universit\'e Montpellier, Montpellier, France}

\date{\today}



\begin{abstract}
The emergence of particle irreversibility in periodically driven 
colloidal suspensions has been interpreted as resulting either 
from a nonequilibrium phase transition to an absorbing state or
from the chaotic nature of particle trajectories. 
Using a simple model of a driven suspension 
we show that a nonequilibrium phase transition
is accompanied by hyperuniform static density fluctuations
in the vicinity of the transition, where we also observe strong 
dynamic heterogeneities reminiscent of dynamics in glassy materials.
We find that single particle dynamics becomes intermittent and strongly 
non-Fickian, and that collective dynamics becomes spatially correlated over 
diverging lengthscales. Our results suggest that the two theoretical scenarii 
can be experimentally discriminated using particle-resolved 
measurements of standard static and dynamic observables. 
\end{abstract} 

\maketitle

Nonequilibrium phase transitions have been studied intensively 
in recent years~\cite{hinrichsen,lubeck-review}. 
Whereas many theoretical models have been 
analysed and organized in a small number of universality classes
(such as directed percolation), 
convincing experimental realisations have typically proved harder to achieve.   
Non-Brownian colloidal suspensions (such as stabilized droplet emulsions or
large particles suspended in a viscous solvent) driven by 
a low-frequency periodic shear flow represent one 
potential realisation of a second-order 
phase transition towards an absorbing 
state~\cite{Pine-Nat,Pine-Nat-Phys,corte-soc,gollub,keim1,keim2,guazzelli}.
It has been found experimentally that below a certain shearing amplitude
(which depends on the density), 
the system evolves after a transient 
into a reversible state where all particles return to 
the same position at the end of each cycle of the 
periodic drive. Above a well-defined threshold amplitude, 
particle motion are no longer periodic, and a continuous 
increase of diffusive motion is observed in this irreversible 
phase~\cite{Pine-Nat}. 

Several
studies~\cite{Pine-Nat-Phys,corte-soc,gollub,ramaswamy,reichhardt,reichhardt2,ganapathy} 
suggested that the experimental transition is in the universality class 
of directed percolation (or conserved directed percolation). This 
interpretation is further supported by an elegant numerical
model of the original experiment, which was shown to 
undergo a second-order nonequilibrium phase transition~\cite{Pine-Nat-Phys}.
However, an alternative explanation
was also proposed~\cite{bartolo-kurchan,butler1,bartolo-nat-comms,reichhardt3,stroock,butler}, 
which relies on the chaotic nature of trajectories in dynamical systems.
In this view, a phase transition is not needed to explain 
the relatively sharp onset of irreversilibity observed in the experiments.
Experiments have not fully established criticality
because direct measurements of the critical exponents are 
difficult~\cite{Pine-Nat,Pine-Nat-Phys,bartolo-nat-comms,ganapathy}.
As a result, the nature of the initial experimental observations remains 
to be fully understood. Here we establish that measurements 
based on standard particle-resolved observables developed in the context
of glassy dynamics~\cite{berthier-book} very directly 
reveal nonequilibrium criticality, when present. 
This suggests that the two existing theoretical scenarii can be 
experimentally discrimated using standard static and dynamic observables.  

\begin{figure}[b]
\begin{centering}
\includegraphics[width=8.5cm]{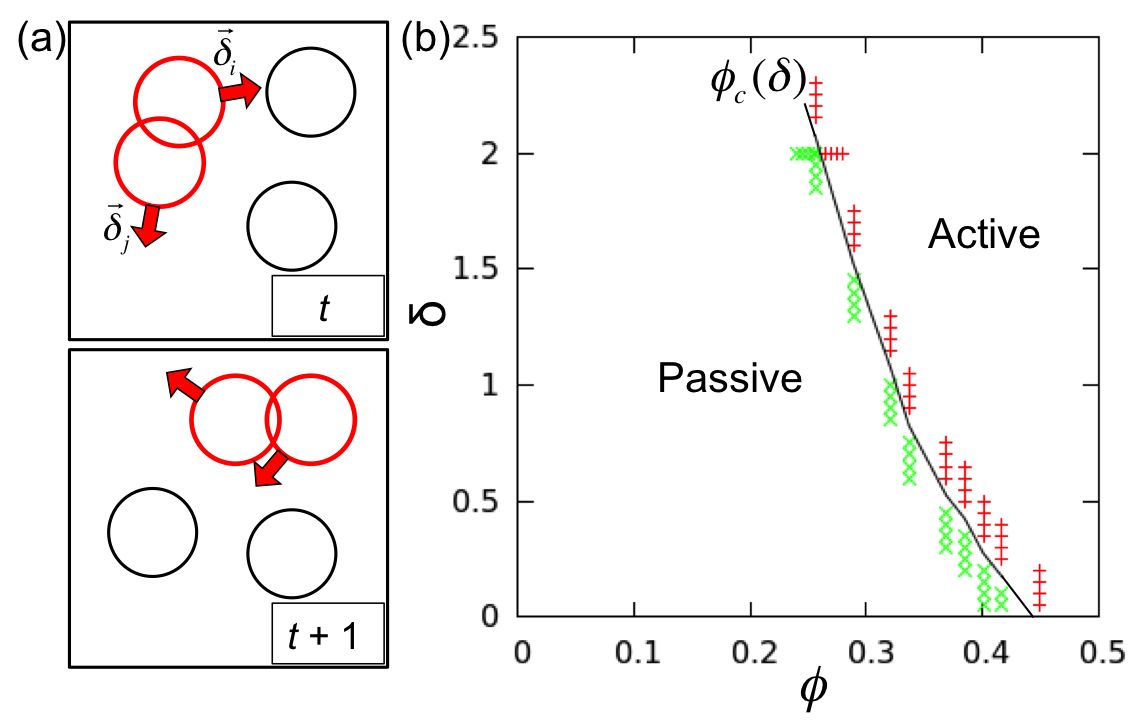}
\par\end{centering}
\protect\caption{
(a) Sketch of the model:
Particles overlaping at time $t$ (red) are simultaneously moved by an
independent random amount. Particles with no overlap (black) are 
immobile, but may become mobile at later time (color online). 
(b) The $(\phi,\delta)$ phase diagram with
a passive region where the number of mobile 
particles vanishes at long time and 
an active phase where particle overlaps are constantly 
destroyed and created. The line of second-order critical points 
is determined from investigating the state points shown with symbols.
\label{fig:model}}
\end{figure}

To support our conclusions, we consider a modified version of the 
model proposed in Ref.~\cite{Pine-Nat-Phys}, as illustrated 
in Fig.~\ref{fig:model}(a). We consider a bidimensional assembly of 
spherical particles of diameter $\sigma$, using periodic boundary 
conditions in a box of linear size $L$. 
The system is initiated from a random configuration, where 
particle overlaps may exist. At each time step, we 
simultaneously move all particles 
which overlap with one neighbor (or several) by an independent random amount.
The displacement of particle $i$ is of the form
$\vec \delta_i = \epsilon_i \hat{e}_i$, 
where $\hat{e}_i$ is a unit vector whose orientation is uniformly 
distributed on a unit circle and the magnitude $\epsilon_i$ is  
uniformly distributed on the interval $[0,\delta]$.
The time is then incremented by one unit.
The model has two control parameters:  
the area fraction $\phi=\frac{\pi N \sigma^2}{4L^2}$, and the  
maximal amplitude of the random kicks $\delta$. We use $\sigma$
as the unit length and we vary 
the area fraction by changing the number of particles $N$ while 
keeping the system size fixed at $L=280$ (unless mentioned otherwise). 

Our model represents an isotropic version of the periodically 
sheared system considered in Ref.~\cite{Pine-Nat-Phys}, where random 
kicks were given to particles virtually colliding with neighbors
during a shear deformation cycle (the shear cycle 
is actually not performed). 
This original rule is in fact equivalent to giving
a random kick to each particle having at least one neighbor 
in an anisotropic area near its center~\cite{schmiedeberg}. 
In our model, we consider that this area is circular, and $\sigma$ 
represents its diameter. 
This small simplification makes the determination of the 
critical properties of the model simpler because it prevents  
the development of locally anisotropic correlations~\cite{schall},
which could affect the numerical value of the 
measured exponents, but not the overall qualitative behavior that 
we report.    
Our setup is also physically meaningful, as it depicts 
the experimental situation where a non-Brownian colloidal 
suspension is driven periodically by a periodic change of the particle 
diameters leading to irreversible collisions. This is obviously
equivalent to isotropic compression cycles 
of a colloidal system. Such experiments could be realized 
experimentally using thermosensitive colloidal particles~\cite{ganapathy}.
 
As expected~\cite{Pine-Nat-Phys} we find that below a
critical density $\phi_c(\delta)$, 
the number of active particles evolves to zero (no more 
overlap) and all particles stop moving; this corresponds to the 
``reversible'' phase of the experiments with periodic forcing.
Above $\phi_c(\delta)$,
the number of active particles fluctuates at steady state around 
its mean non-zero value, and the system is diffusive. 
By carefully exploring the steady state 
properties~\cite{footnote} 
of the state points shown in Fig.~\ref{fig:model}(b), we have 
numerically determined the critical line 
$\phi_c(\delta)$ separating the two phases.
To determine the critical properties of the model, we used
the order parameter, which is the fraction of 
active particles, $f_a(t) = N_a / N$, where $N_a(t)$ is the number of particles
having overlaps at time $t$. The spatio-temporal properties
of $f_a(t)$ display critical properties that can be compared to known 
universality classes~\cite{elsen}. 
While such a study is not problematic for 
computer simulations, it is more difficult in experiments as it 
requires tracking the displacement of all particles at all times and separating 
mobile from immobile particles. We provide below simpler observables  
which exhibit relevant signatures of the underlying phase transition.

A simpler quantity, measured in the original experimental
study, is the single particle diffusion constant, 
defined as $D = \lim_{t \to \infty} \langle |\Delta \vec{r}(t)|^2 \rangle/(4t)$, 
where $\Delta \vec{r}(t)$ represents the displacement of a 
given particle over a time $t$. The brackets indicate 
an ensemble average (equivalent, in steady state, to a time average).
This measurement can be performed using tracer particles
followed over long times.    
In our model, we find that $D=0$ for $\phi<\phi_c$, and it emerges 
continuously above $\phi_c$: 
\begin{equation}
D \sim (\phi-\phi_c)^\beta, \quad \phi\rightarrow\phi_c^+.
\label{diffusion}
\end{equation}
We measure $\beta \simeq 0.572$, 
which is similar to the value found in related 
studies~\cite{Pine-Nat-Phys,reichhardt,reichhardt2,ganapathy,foffi}.
Additionally, we find that $\beta$ remains constant,
within statistical uncertainty,  
along the line $\phi_c(\delta)$. For most of this paper, 
we thus fix $\delta=0.5$ for which 
$\phi_c \simeq 0.375$. 
The critical exponent $\beta$ in Eq.~(\ref{diffusion}) is 
relevant because it is directly related  to the order parameter, 
$\langle f_a \rangle$.  
To see this, let us rewrite the particle displacement as  
$\Delta \vec{r}(t) = \sum_{t'=0}^{t-1} [\vec{r}(t'+1)-\vec{r}(t')]$. Denoting
by $t_a$ the number of timesteps where the tracer is mobile between
times $0$ and $t$, the displacement is the sum of $t_a$ random kicks.
As a result, $D$ scales as $t_a/t$, which represents the fraction of 
the time when the tracer is mobile. When $t \to \infty$, 
this becomes the ensemble average $\langle f_a \rangle$. 
In two dimensions, 
$\beta \approx 0.58$ for directed percolation, and $\beta \approx 
0.64$ for conserved directed percolation (or `Manna' universality 
class)~\cite{lubeck-review}. Our simulations appear
closer to the directed percolation universality class. 

\begin{figure}
\includegraphics[width=8.5cm]{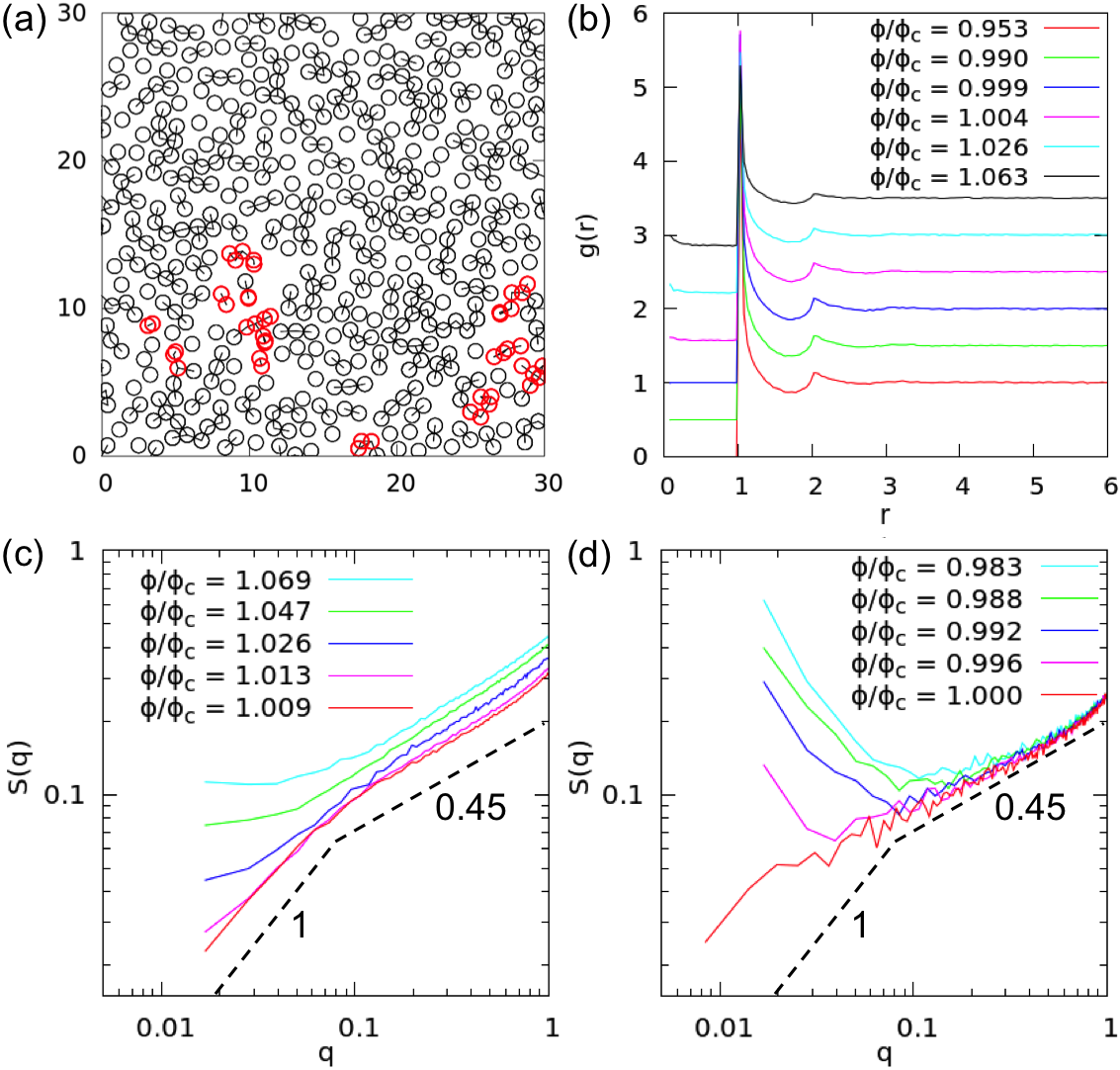}
\caption{
(a) Typical snapshot close to criticality 
($\phi/\phi_c=1.004$ and $\delta=0.05$) at steady state. 
Active/passive particles are shown in red/black. 
Bonds are drawn between particles whose separation is less than 
$1.05$ to reveal string-like clusters. 
(b) The radial distribution function for different 
densities (for $\delta =0.05$) reveals two peaks at $r=1$ and $r =2$, 
due to the strings but $g(r)$ does not change 
significantly across the transition.
(c) The structure factor for different densities $\phi>\phi_c$
reveals hyperuniformity at large scale close to criticality 
$\phi_c\simeq 0.37499$, with $S(q \to 0)\sim q$
crossing over to a different power law.
(d) The structure factor for $\phi<\phi_c$
behaves similarly as $\phi_c$ is approached.
The system size is $L=280$ except for $\phi/\phi_c=1.000$ for 
which $L=560$ is used. 
\label{fig:hyperuniform}}
\end{figure}

\begin{figure*}
\begin{centering}
\includegraphics[width=15.cm]{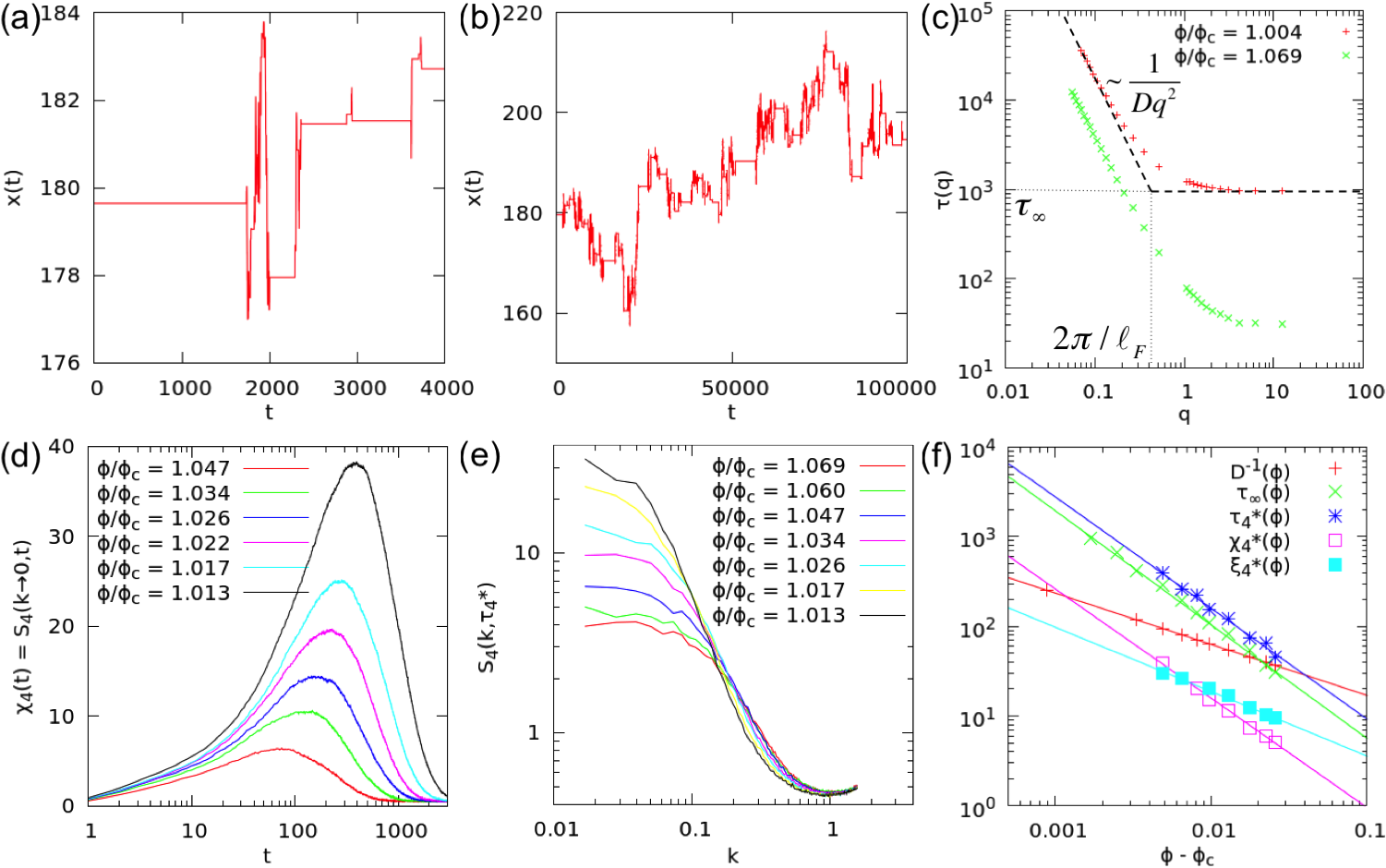}
\par\end{centering}
\protect\caption{
(a) Typical particle trajectory along the $x$-axis 
for $\phi/\phi_c=1.009$
characterised by long waiting periods (when the particle is passive)
and few jumps (when the particle is active).
(b) Over a much longer time interval, 
the same trajectory resembles an ordinary random walk.
(c) Wavevector dependence of the relaxation 
timescale $\tau(q)$ for two different densities. A non-Fickian 
($\tau \sim \tau_{\infty}(\phi)$) to Fickian ($\tau \sim 1/(Dq^{2})$) 
crossover is observed at a wavector ($2\pi/\ell_F(\phi)$) which decreases 
as the transition is approached.
(d) The four-point susceptibility $\chi_{4}(t)$ quantifies spatially correlated
dynamics over a time interval $t$. It shows a peak at $t=\tau_{4}^{*}(\phi)$
which diverges as $\phi \to \phi_c^+$.  
(e) Four-point structure factor $S_4(k,t=\tau_4^*)$ as a function of the 
wavevector $k$ for different densities reveals 
a diverging dynamic correlation length $\xi_4^*(\phi)$.
(f) Critical power laws for quantities measured in this work:
inverse diffusion constant $D^{-1}$ (exponent 0.572),
Fickian crossover timescale $\tau_\infty$ (1.27), 
dynamic timescale $\tau_4^*$ (1.24), maximum susceptibility $\chi_4^*$ (1.22), 
and dynamic lengthscale $\xi_4^*$ (0.72).
\label{fig:tau-q}}
\end{figure*}

In experiments with non-Brownian particles, it is easy to visualise 
particle configurations and analyse static density fluctuations. 
In Fig.~\ref{fig:hyperuniform}(a),
we show a snapshot of the system close to criticality, where very 
few active particles coexist with many passive ones. The structure
appears globally homogeneous with no sign of large scale density 
fluctuations. At smaller scale, particles form short one-dimensional
clusters, or `strings', which are disconnected and do not percolate 
throughout the system. This tendency is confirmed in
the radial distribution function~\cite{hansen}, $g(r)$, shown in 
Fig.~\ref{fig:hyperuniform}(b) for various densities across 
the critical point. We see that $g(r)$ has two peaks at $r=1$ and 
$r= 2$, indicative of the string-like structure at short lengthscales.
The sharpness of these two peaks is controlled by the amplitude $\delta$
of the random jumps; they become sharper as $\delta \to 0$.  
Very similar radial distribution functions
have recently been observed in a periodically driven 
colloidal suspension~\cite{bartolo-nat-comms}. 
More importantly, we conclude from 
Fig.~\ref{fig:hyperuniform}(b) that $g(r)$ is rather
insensitive to the crossing of the phase transition. 

While this might indicate that static density fluctuations 
are insensitive to the critical point, Fourier transforming 
$g(r)$ to get the structure factor $S(q)$ shows interesting 
behaviour, as suggested very recently~\cite{levine}. 
In Figs.~\ref{fig:hyperuniform}(c,d) we show the low-$q$ behavior of 
$S(q)$ respectively above and below the critical point. In this log-log
representation it is clear that $S(q \to 0)$ becomes extremely small 
as $\phi \to \phi_c^{\pm}$, with emerging power laws.
Notice that $S(q)$ converges to the same form on both sides, 
but convergence from the absorbing phase is slower, as the system retains
memory of the disordered initial conditions on very large scale. 
However, careful analysis of the density 
evolution~\cite{elsen} reveals that $S(q \to 0)$ vanishes
precisely at $\phi_c$, where $S(q) \approx q$ up to $q \approx 0.05$,
crossing over to $S(q) \approx q^{0.45}$ at larger $q$.  
A vanishing $S(q \to 0)$ physically means that density 
fluctuations are strongly suppressed at large scale, which 
is termed `hyperuniformity'~\cite{stillinger-review}. 
The linear behaviour with $q$ implies 
that the number of particles in a (large) subsystem of size $L$
obeys $\langle \Delta N^2 \rangle / \langle N \rangle \sim L^{-1}$.
In an equilibrium fluid, this ratio is instead
independent of $L$. Hyperuniformity 
has been reported in a number of nonequilibrium 
situations~\cite{stillinger-review,stillinger-prl,berthier-hyperuniform,bird,rob}, among which hard sphere jammed packings. 
However the critical density here is much smaller than the 
jamming density and the hyperuniform structure is different from 
that of compressed hard spheres. 
A previous study~\cite{levine} suggested that 
$S(q) \sim q^{0.45}$ reflects the asymptotic behavior of $S(q)$, 
whereas we find that this is only a transient. 
These findings imply nonetheless that 
static fluctuations reveal a striking signature of 
criticality, which has not yet been investigated 
experimentally~\cite{Pine-Nat,ganapathy,bartolo-nat-comms}. 

We now turn to the dynamics. Close to the irreversibility transition, 
we detect strong signatures of dynamic 
heterogeneities, reminiscent of observations in 
disordered systems approaching dynamic arrest (such as dense 
colloidal suspensions)~\cite{berthier-book}. This analogy is useful, as 
it provides us with a toolbox to directly reveal 
the criticality associated to the nonequilibrium phase transition. 

A striking observation stems from tracer trajectories, 
see Figs.~\ref{fig:tau-q}(a,b).
Over short times, Fig.~\ref{fig:tau-q}(a), the trajectory 
is characterised by long waiting periods (when 
the particle is passive)
and a few moments where the particle makes several rapid jumps 
(when the particle is active).
Because activity is sparse close to $\phi_c$,
particles are necessarily immobile most of the time.
Such intermittency is also observed in 
glassy fluids where particles are caged over long 
periods~\cite{berthier-fickian}. 
Much longer trajectories resemble ordinary Brownian motion,  
Fig.~\ref{fig:tau-q}(b), suggesting that Fickian diffusion 
is recovered at large scale. 
Intermittent, non-Fickian dynamics thus represents another 
signature of the criticality, which could be systematically
investigated experimentally through
the self-intermediate function $F_s(q,t)= 
\left\langle \frac{1}{N} \sum_{i} c_i(q,t) \right\rangle$,
where $c_i(q,t) = \cos \left[ \vec{q} \cdot \Delta \vec{r}_i(t) \right]$.
Physically, $F_s(q,t)$ relaxes from 1 to 0 when particles
have moved an average distance $\frac{2 \pi}{q}$. 

The relaxation time $\tau(q)$ (defined as $F_s(q,\tau)=e^{-1}$)
is plotted in Fig.~\ref{fig:tau-q}(c).
Over large distances ($q \to 0$), 
Fickian behaviour is observed, $\tau(q) \sim 1/(Dq^2)$.
On the other hand, at shorter lengthscales 
$\tau(q)$ crosses over to a plateau 
value, $\tau_\infty(\phi)$. As $\phi \to \phi_c^+$, 
this non-Fickian plateau regime becomes dominant. 
Physically, $\tau_\infty$ represents the typical waiting time 
before an immobile particle becomes active. We measure   
$\tau_\infty\sim(\phi-\phi_c)^{-\nu_\parallel}$,
where $\nu_\parallel \simeq 1.27$, see Fig~\ref{fig:tau-q}(f).
Interestingly, we found numerically that the same exponent 
$\nu_\parallel$
controls the temporal fluctuations of 
the order parameter, $f_a(t)$~\cite{elsen}. This 
is close to the directed percolation value $\nu_\parallel = 1.30$
($\nu_\parallel = 1.23$ for conserved directed 
percolation)~\cite{lubeck-review}.  Finally,
because $\tau_\infty$ and $D^{-1}$ obey different power laws, 
we can define a diverging crossover 
lengthscale for the emergence of Fickian diffusion~\cite{berthier-fickian}, 
$\ell_F \sim \sqrt{D \tau_\infty} \sim (\phi - \phi_c)^{-(\nu_\parallel - 
\beta)/2}$ (see Fig.~\ref{fig:tau-q}(c)),
indicating that diffusion is non-Fickian at all length scale 
at the critical point.

Intermittency suggests that mobile and immobile 
particles coexist in space. We now show that the associated  
dynamic fluctuations also diverge at $\phi_c$. 
To this end, we study spatial correlations of particle displacements, 
in analogy with measurements in dense fluids~\cite{berthier-book}.
We first introduce a `four-point' structure factor 
$S_4(k,t)$~\cite{berthier-book}:
\begin{equation}
S_4(k,t) = \left\langle \frac{1}{N}\sum_{i,j}  
e^{i\mathbf{k}\cdot(\mathbf{r}_i-\mathbf{r}_j)} c_i(q,t) c_j(q,t)) \right\rangle,
\end{equation}
where we fix $q=2\pi$ [in the plateau regime 
of $\tau(q)$ in Fig~\ref{fig:tau-q}(c)].
Physically, $S_4(k,t)$ measures (in the Fourier domain) spatial 
correlations between particles which have moved a distance $q^{-1}$ 
during the time interval $t$. We also define the 
four-point susceptibility $\chi_4(t) = S_4(k \rightarrow 0, t)$,
which measures the variance of spontaneous fluctuations of the time correlation 
function $F_s(q,t)$.
 
The dynamic susceptibility is plotted in Fig.~\ref{fig:tau-q}(d) for 
different densities. For a given $\phi$, $\chi_4(t)$
exhibits a maximum, $\chi_4^*$, at a time $\tau_4^*$.
Both $\tau_4^*$ and $\chi_4^*$ grow rapidly as $\phi_c$ is 
approached, and obey power laws, see Fig.~\ref{fig:tau-q}(f):
$\tau_4^*\sim(\phi-\phi_c)^{-\nu_\parallel}$, with 
$\nu_\parallel\simeq1.24$, compatible with the 
result for  $\tau_\infty$.
Similarly, $\chi_4^* \sim (\phi-\phi_c)^{-\gamma}$, with  
$\gamma \simeq 1.22$. 
The divergence of $\chi_4^*$ is accompanied by a 
diverging correlation length, 
as revealed by the evolution of $S_4(k,t)$ in Fig.~\ref{fig:tau-q}(e).
Here we fix $t=\tau_4^*$ when the correlation is maximal. 
We observe a growing peak at low wavevector shifting to lower
$k$ as $\phi$ increases. 
We follow established procedures~\cite{berthier-book} 
and extract the dynamic lengthscale 
$\xi_4^*$ by using the following scaling form:
$S_4(k,\tau_4^*) / \chi_4^* = F(k\xi_4^*(\phi))$,
where $F(x)$ is a scaling function independent of $\phi$.
As shown in Fig.~\ref{fig:tau-q}(f), $\xi_4^*$ 
obeys a power law divergence, $\xi_4^*\sim(\phi-\phi_c)^{\nu_\perp}$.
We measure $\nu_\perp\simeq0.72$. We found numerically 
that a similar critical exponent controls the divergence of the 
order parameter correlation length~\cite{elsen}. Again, our measurements
compare well to the directed 
percolation exponent, $\nu_\perp = 0.72$ ($\nu_\perp = 0.80$ for 
conserved directed percolation)~\cite{lubeck-review}.  

Our results demonstrate that the irreversibility transition 
observed in periodically driven systems has 
interesting qualitative analogies with glassy systems. In both cases, 
the radial distribution function $g(r)$ appears insensitive to  
dynamic arrest, whereas other quantities display stronger signatures. 
We have reported a strong suppression of the density fluctuations 
at large scales, and a divergence of several dynamic quantities
associated to single particle and collective dynamics. 
The analogy between the two types of systems 
suggests that particle-based measurements and observables 
developed for glassy materials could prove useful in driven suspensions.  
These tools could in particular reveal whether the 
`singularity-free' explanation based on the Lyapunov instability 
is experimentally relevant. Our work also suggests that it would 
be interesting to characterize
more precisely the static structure and dynamic correlations 
in the vicinity of the yielding transition in dense suspensions 
under oscillatory shear~\cite{keim1,ganapathy,reichhardt3,foffi,luca}, 
which represents another important situation where a reversibility 
transition and a transition to chaos \cite{chaos} should be 
better understood.

\acknowledgments
We thank D. Bartolo and R. Jack for useful discussions. 
The research leading to these results has received funding from the European
Research Council under the European Union’s (EU) Seventh
Framework Programme (FP7/2007-2013)/ERC Grant Agreement No. 306845.


\begin{thebibliography}{99}
\bibitem{footnote}
For a given $\phi<\phi_c$, we run several simulations from different 
initial random configurations 
and wait until each simulation reaches an absorbing state.
Static quantities are then ensemble-averaged over different runs.
On the other hand for $\phi>\phi_c$, long time average is sufficient 
since the system is ergodic.
\end{thebibliography}
\end{document}